# *Ab initio* simulations on the pure Cr lattice stability at 0K: Verification with the Fe-Cr and Ni-Cr binary systems

Songge Yang[1], Yi Wang[2], Zi-Kui Liu[2], Yu Zhong[1]

[1]Department of Material Science and Engineering, Worcester Polytechnic Institute, 100 Institute Rd, Worcester, MA, 01609, USA

[2]Department of Materials Science and Engineering, Pennsylvania State University, University Park, PA 16802, USA

## Abstract:

Significant discrepancies have been observed and discussed on the lattice stability of Cr between the predictions from the *ab initio* calculations and the CALPHAD approach. In the current work, we carefully examined the possible structures for pure Cr and reviewed the history back from how Kaufman originally determined the Gibbs energy of FCC-Cr in the 1970s. The reliability of Cr lattice stability derived by the CALPHAD and *ab initio* approaches was systematically discussed. It is concluded that the Cr lattice stability based on the CALPHAD approach has large uncertainty. Meanwhile, we cannot claim that the *ab initio* $H_{\text{FCC-Cr}}$ is error-free as FCC-Cr is an unstable phase under ambient conditions. The present work shows that the *ab initio* $H_{\text{FCC-Cr}}$ can be a viable scientific approach. As both approaches have their limitations, the present work propose to integrate the *ab initio* results into the CALPHAD platform for the development of the next generation CALPHAD database. The Fe-Cr and Ni-Cr binary systems were chosen as two case studies demonstrating the capability to adopt the *ab initio* Cr lattice stability directly into the current CALPHAD database framework.

**Keywords**: *Ab initio*, CALPHAD, lattice stability, Cr

# 1. Introduction

Lattice stability, which was originally proposed by Kaufman [1], is defined as the Gibbs energy difference for a pure element in two different states. It is an essential quantity in the Computer Coupling of Phase Diagrams and Thermochemistry (CALPHAD) approach to construct Gibbs energy of binary, ternary, and higher-order systems.

Currently, the lattice stability can be derived with two approaches. The first one is the CALPHAD approach developed by Scientific Group Thermodata Europe (SGTE). In this approach, the lattice stability was obtained semi-empirically through the extrapolations in either temperature-pressure or temperature-composition phase diagrams. The second one is the *ab initio* approach, which predicts the energy difference of two crystal structures by solving the Schrödinger equation for the electron density. Since the 1970s, researchers started to investigate the lattice stabilities of pure elements developed by these two approaches [1, 2] and debated since the early days [3]. In most cases, the lattice stability obtained by these two approaches agrees fairly well with each other, the differences are within 5kJ/mol. However, there are some notable exceptions, such as FCC-Cr, Mo, and W, showing significant discrepancies by approximately 20-30kJ/mol [4]. In 2004, Wang et al. [5] systematically investigated the lattice stabilities of 78 pure elements and compared their differences between *ab initio* calculations and SGTE. They observed these discrepancies are less than 10kJ/mol for Co, Ti, Zr, and Mn. However, the discrepancies are larger than 20kJ/mol for Cr, Mo, W, and Ru, and reach 55kJ/mol for Os. It has been reported that most of these remarkable discrepancies are observed in the 3d, 4d, and 5d transition elements, especially for the unstable structures [5-8]. In the present paper, the stable, metastable and unstable structures are defined by the second derivatives of Gibbs energy with respect to its natural variables [9], i.e. positive for a stable system, zero for the limit of stability and negative for an unstable system [10].

More recently, the "*Inflection detection method*" [11, 12] and "*ab initio molecular dynamic (AIMD) method*" [13] have been proposed to reduce the discrepancies and achieved values which are closer to SGTE. The essential question remains how to correctly define and evaluate the lattice stability of the mechanically unstable structures, since the reliability of the pure elements lattice stabilities will greatly affect the extrapolations to the high-order systems [14, 15]. In the present study, the emphasis is placed primarily on investigating and discussing the reliability of Cr lattice stability between the *ab initio* and CALPHAD approaches. The Fe-Cr and Ni-Cr binary systems were chosen as two case studies to test if the *ab initio* Cr lattice stability can be directly used in the CALPHAD modeling.

# 2. Methodology

## 2.1 Ab initio computational details

The *ab initio* calculations were performed by using the Vienna ab initio simulation package (VASP) [16, 17]. The generalized gradient approximation (GGA) [18] with the Perdew-Burke-Ernzerhof (PBE) exchange-correlation function [19] was used to describe exchange and correlation effects. The k-point meshes (Monkhorst-pack scheme for cubic, Gamma-centered scheme for hexagonal) with density not less than 5000 pra (per-reciprocal atom) were used to sample the Brillouin zones. The accurate total energy calculations were performed by means of the linear tetrahedron method with Blöchl's corrections [20]. In all cases, the total energies obtained by relaxation calculations were converged to $10^{-6}$ eV/cell with a 1.75 times plan wave cutoff energy suggested by corresponding element pseudopotential. Meanwhile, the static calculations, which are converged to $10^{-8}$ eV/cell with 520eV cutoff-energy, were applied after each relaxation procedures. The initial spin orientations are set to be ferromagnetic for all the pure elements and binary alloys except for FCC-Fe (double layer antiferromagnetic) and BCC-Cr (antiferromagnetic). The Phonon frequency calculations are carried out by the supercell method as implemented in the *YPHON* code [21] with VASP as the computational engine. The 64-atoms FCC-Cr was applied for phonon calculations. The k-point mesh was selected as the $3\times3\times3$ gamma-centered scheme. The forces induced by small displacement were calculated.

## 2.2 Selections of special quasi-random structures (SQSs)

The special quasirandom structures (SQS) approach aims to find the small-unit-cell periodic structures with $\prod_{k,m}(SQS) \cong \prod_{k,m}(R)$ for as many figures as possible [22-24], where $\prod_{k,m}(R)$ is the correlation function of a random alloy, which is simply shown by $(2x-1)^k$ in a $A_{1-x}B_x$ substitutional binary alloy, in which $x$ is the composition of the alloy, and $k$ is the number of atoms (pair, triplet, quadruplets) considered in the correlation functions. The structures exhibiting the less correlation function mismatch will be selected for the *ab initio* calculations since these structures are closer to random alloys. In the present work, the 16-atoms and 32-atoms FCC and BCC SQSs supercells were generated by the alloy-theoretic automated toolkit (ATAT) package [25] with *mcsqs* code [26]. The range of pairs, triplets and quadruplets are selected as 2.0, 1.6 and 1.0, respectively (-2=2.0, -3=1.6, -4=1.0) with *corrdump* command [25]. The configurations of the SQS structures are shown in Figure 1. Meanwhile, the correlation mismatch of generated SQS supercells is shown in the *Supplementary file*.

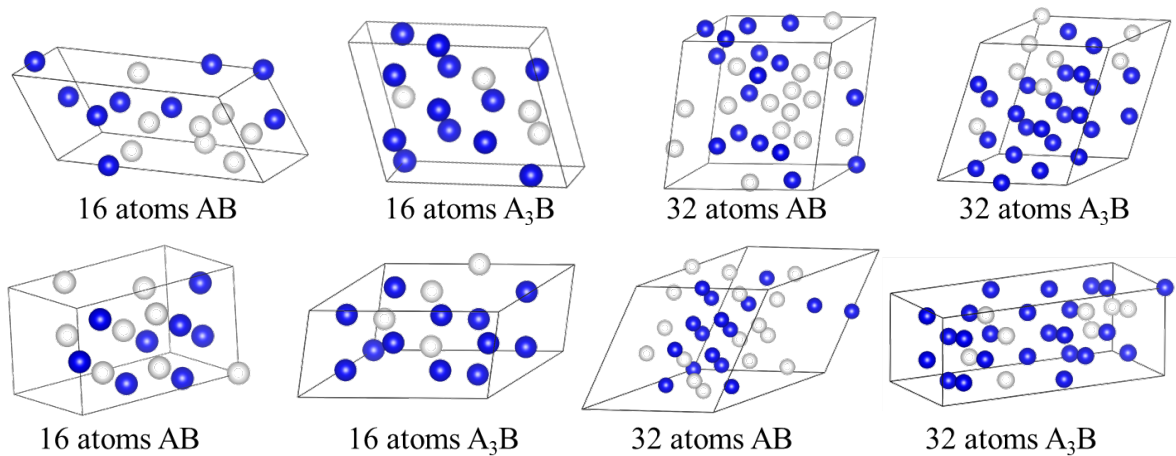

16 atoms AB     16 atoms A₃B     32 atoms AB     32 atoms A₃B

16 atoms AB     16 atoms A₃B     32 atoms AB     32 atoms A₃B

*Figure 1. Selected SQS structures for ab initio calculations (a)-(d) BCC, (e)-(h)FCC*

## 2.3 Thermodynamic modeling

The FCC phase in the Fe-Cr and Ni-Cr binaries are modeled in the current work. The Gibbs free energy per mole of atoms $G_m^{FCC}$ of Fe-Cr (Eq. (1)) and Ni-Cr (Eq. (2)) can be described by a substitutional solution model:

$$G_m^{FCC} = x_{Cr}\,^0G_{Cr}^{FCC} + x_{Fe}\,^0G_{Fe}^{FCC} + RT(x_{Cr}lnx_{Cr} + x_{Fe}lnx_{Fe}) + {}^{xs}\Delta G_m^{FCC} + {}^{mag}G_m^{FCC} \qquad (1)$$

$$G_m^{FCC} = x_{Cr}\,^0G_{Cr}^{FCC} + x_{Ni}\,^0G_{Ni}^{FCC} + RT(x_{Cr}lnx_{Cr} + x_{Ni}lnx_{Ni}) + {}^{xs}\Delta G_m^{FCC} + {}^{mag}G_m^{FCC} \qquad (2)$$

where $^0G$ is the molar Gibbs free energy of pure element, $^{mag}G$ represents the magnetic contribution of Gibbs free energy, and $^{xs}\Delta G$ represents the excess Gibbs free energy, which can be expressed by Redlich-Kister polynomial [27] as:

$${}^{xs}G_m^{FCC} = x_{Cr}x_{Fe}\sum_{j=0}^{n} {}^jL_{Cr,Fe}^{FCC}\,(x_{Cr} - x_{Fe})^j \qquad (3)$$

$${}^{xs}G_m^{FCC} = x_{Cr}x_{Ni}\sum_{j=0}^{n} {}^jL_{Cr,Ni}^{FCC}\,(x_{Cr} - x_{Ni})^j \qquad (4)$$

where $^jL_{Cr,Fe}^{FCC}$ and $^jL_{Cr,Ni}^{FCC}$ are the $j$th binary interaction parameters of Fe-Cr and Ni-Cr binary systems, respectively, which can be expressed as $^jL_{Cr,Fe}^{FCC} = {}^jA_{Cr,Fe}^{FCC} + {}^jB_{Cr,Fe}^{FCC}T$ and $^jL_{Cr,Ni}^{FCC} = {}^jA_{Cr,Ni}^{FCC} + {}^jB_{Cr,Ni}^{FCC}T$.

For the pure elements, $^0G_{Cr}^{FCC}$, $^0G_{Fe}^{FCC}$, and $^0G_{Ni}^{FCC}$ describe the FCC Gibbs free energy of Cr, Fe, and Ni. They are related to the stable forms as:

$${}^0G_{Cr}^{FCC} - {}^0G_{Cr}^{BCC} = A_{Cr} - B_{Cr}T \qquad (5)$$

$${}^0G_{Fe}^{FCC} - {}^0G_{Fe}^{BCC} = A_{Fe} - B_{Fe}T \qquad (6)$$

The linear expressions on Eqs. (3) and (4) are referred to as the "*lattice stability*" in accordance with the CALPHAD convention. The constant $A_i$ and $B_i$ represent the enthalpy and entropy differences between the FCC and BCC phases of the elements $i$. Since the stable state of pure Ni is FCC structure, this value is equal to zero.

The discrepancy between the *ab initio* and CALPHAD lattice stability is neglectable for Fe and Ni, but significantly large for Cr [5, 12, 28-31]. In the present work, the total energy of FCC-Cr from the *ab initio* calculation was adopted to reproduce the Fe-Cr and Ni-Cr phase diagram by tuning the FCC interaction parameters.

# 3. Results

## 3.1 Total energy calculations of Cr

The calculated static energies of FCC, BCC and HCP Cr as a function of volume, together with the fittings by seven equations of states (EOS's) are shown in Figure 2. The widely used linear EOS's are the Birch-Murnaghan EOS's [32], which can be calculated and fitted by:

$$E(V) = a + \frac{B_0 V}{B_0'}\left(1 + \frac{\left(\frac{V_0}{V}\right)^{B_0'}}{B_0' - 1}\right) \qquad (7)$$

In Eq. (5), the fitting parameter $a = E_0 - \frac{B_0 V_0}{B_0' - 1}$, and the parameters $V_0$, $E_0$, $B_0$ and $B_0'$ represent the equilibrium volume, total energy, bulk modulus, and its first derivation with respect to pressure, respectively. The EOS's fitted equilibrium properties are shown in Table 1. Meanwhile, the fitting error ($E_{err}$) can be estimated by:

$$E_{err} = \sqrt{\frac{\sum[(E_{fit}-E_{calc})/E_{calc}]^2}{k}} \tag{8}$$

where $E_{fit}$ and $E_{calc}$ represent the fitted and *ab initio* calculated energies, respectively, and $k$ represents the total numbers of the calculated volumes.

Based on Figure *2*, the total energy of FCC, BCC, and HCP Cr are calculated and fitted by EOS's. Regarding FCC and HCP phases, the final optimized magnetic states are nonmagnetic (NM) states no matter what types of magnetic orders as the inputs. It can be concluded that both FCC and HCP Cr are nonmagnetic (NM) phase. However, for BCC structures, it is observed that the antiferromagnetic (AFM) magnetic state has the lowest total energy, which means AFM Cr-BCC is stable at the ground state [33]. Meanwhile, the slight increase in the $V_0$ and the decrease of the $B_0$ are observed for AFM BCC-Cr due to the magnetic phase transition.

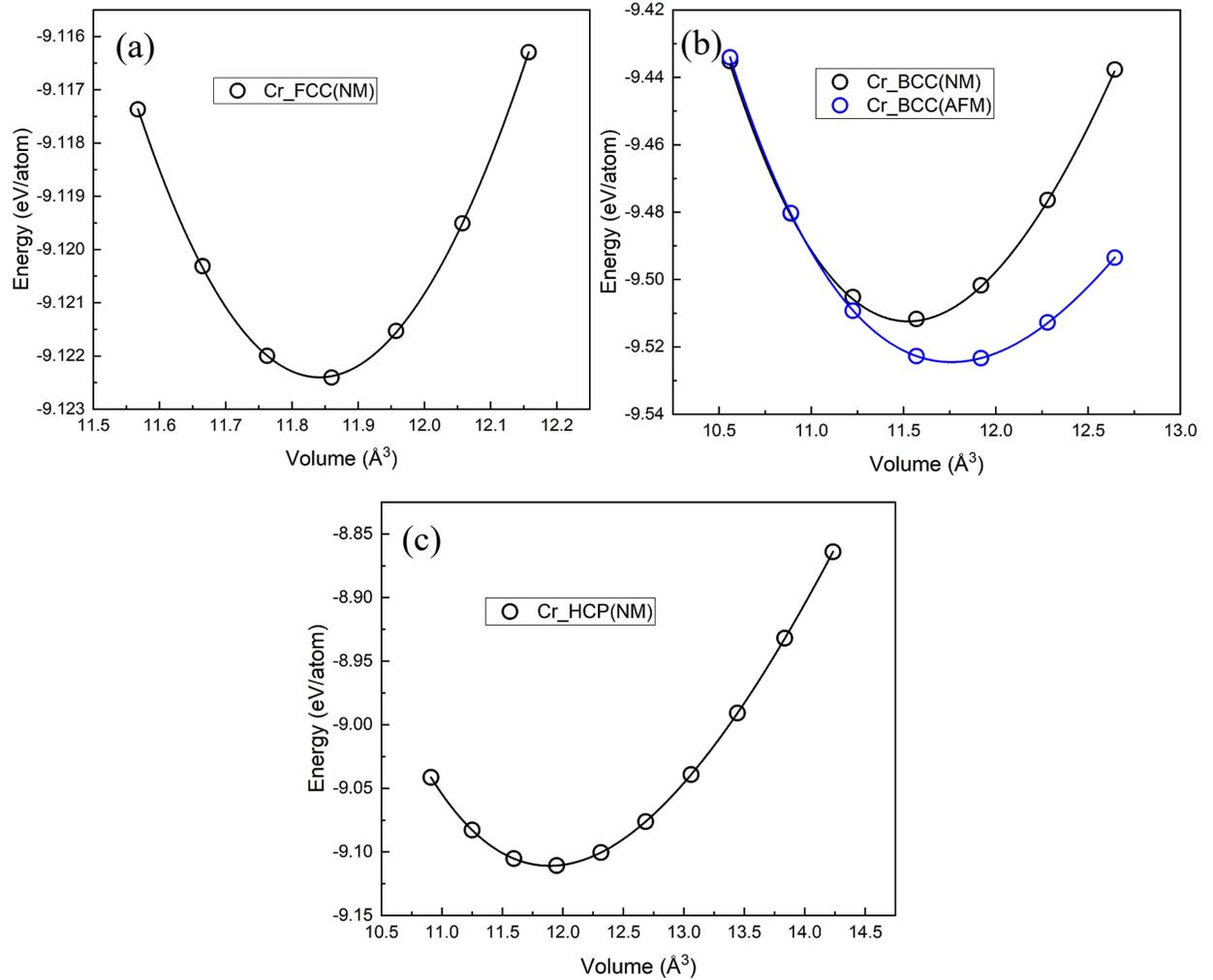

*Figure 2. Calculated total energies of Cr with different structures (a) FCC, (b) BCC, (c) HCP*

*Table 1. The EOS fitting of the ground state calculations of Cr*

| Primitive cell | V0(A$^3$) | E0(eV/atom) | B(V0) (eV/A$^3$) | B'(VO) | Fitting error($10^{-4}$eV/atom) |
|---|---|---|---|---|---|
| Cr-FCC (NM) | 11.84 | -9.1224 | 1.5198 | 4.2966 | 0.3 |
| Cr-BCC (NM) | 11.53 | -9.5119 | 1.6110 | 4.3823 | 0.4 |
| Cr-BCC (AFM) | 11.75 | -9.5245 | 1.1154 | 6.4221 | 1.7 |
| Cr-HCP (NM) | 11.89 | -9.1109 | 1.4679 | 4.1994 | 0.2 |

*3.2 Ab initio calculations of the Fe-Cr and Ni-Cr binary systems*

In the current section, we systematically investigated the enthalpy of mixing ($\Delta H_{mix}$) and enthalpy of formation ($\Delta H_f$) of FCC and BCC phase of Fe-Cr and Ni-Cr binary systems. The $\Delta H_{mix}$ of a stable or metastable phase $\Phi$ can be approximated by Eq. (7) [24]:

$$\Delta H_{mix}^{\Phi}(A_{1-x}B_x) = E^{\Phi}(A_{1-x}B_x) - (1-x)E^{\Phi}(A) - xE^{\Phi}(B) \qquad (9)$$

where $E^{\Phi}(i)$ are the *ab initio* total energy of $A_{1-x}B_x$ and pure elements $A$ and $B$ in the structure $\Phi$. In the current work, the SQS method is adopted for various FCC and BCC binary compositions. Typically, the quality of SQS structures has a great impact on the total energies. The randomness of the SQS structure is not only related to the correlation mismatch but also related to the number of atoms in the supercells. It is much easier to generate a random structure with a larger supercell than a smaller one. Accordingly, considering the quality of the SQS supercells caused by the cell size, both 16 and 32 atoms FCC and BCC SQSs are calculated in the present work. The $\Delta H_{mix}$ of 16-atom (the red circle with solid line) and 32-atom (the blue circle with solid line) SQS BCC supercells in Fe-Cr and Ni-Cr binary systems are plotted in Figure 3, in comparison with values from CALPHAD modeling [34-38] as well as the *ab initio* results from other works [39-41]. The total energy calculations were carried out by fitting the EOS's, and ferromagnetic (FM) magnetic configurations were considered for all the SQS structures. Based on Figure 3, it shows that the $\Delta H_{mix}$ obtained by 16-atom SQS supercells is almost the same as the results calculated from 32-atom SQS supercells in both binary systems, indicating the16-atoms SQSs generated is as random as the 32-atoms SQSs for the BCC structure. As shown in Figure 3 (a), it is observed that the $\Delta H_{mix}$ derived by present work shows good agreement with the CALPHAD results [34, 36-38] and the previous *ab initio* results in the Fe-Cr binary system [42-44]. Meanwhile, our $\Delta H_{mix}$ in the Ni-Cr binary system are slightly lower than the previous *ab initio* prediction proposed by Liu et al. [41], but the present work exhibits relatively better agreement with the CALPHAD results [35]. Specifically, the $\Delta H_{mix}$ difference between our *ab initio* results and CALPHAD is less than 2kJ/mol for Ni$_1$Cr$_1$ and Ni$_1$Cr$_3$, but this difference becomes slightly larger (around 3.8kJ/mol) on Ni-rich side. It is suspected the $\Delta H_{mix}$ from CALPHAD is less reliable when the BCC phase becomes unstable, which contributes to the larger discrepancy on Ni-rich side.

Unlike the BCC phase, significant discrepancies between the *ab initio* and CALPHAD approaches are observed for the FCC phase (Figure 4) in both Fe-Cr (Figure 4 (a)) and Ni-Cr (Figure 4 (b)) binary systems. The $\Delta H_{mix}$ of the FCC phase is negative from the present *ab initio* calculations, while positive from the CALPHAD calculations [34, 37, 38]. The maximum difference is more than 15kJ/mol, which was also observed by Wrobel et al. [44] and Liu et al. [41]. There are two possible contributions to the discrepancies of FCC $\Delta H_{mix}$ between both approaches. First, the FCC phase is metastable on the Fe-rich side and stable on the Ni-rich side, respectively, however it becomes unstable when the composition moves toward the Cr-rich side. It is hard to conclude what the reliable value is for $\Delta H_{mix}$ based on the *ab initio* calculations for any unstable structures. Second, the lattice stability of the end members also plays an important role for such discrepancy. If the pure elements lattice stabilities derived by both approaches have a huge difference, $\Delta H_{mix}$ from both approaches would be very different from each other.

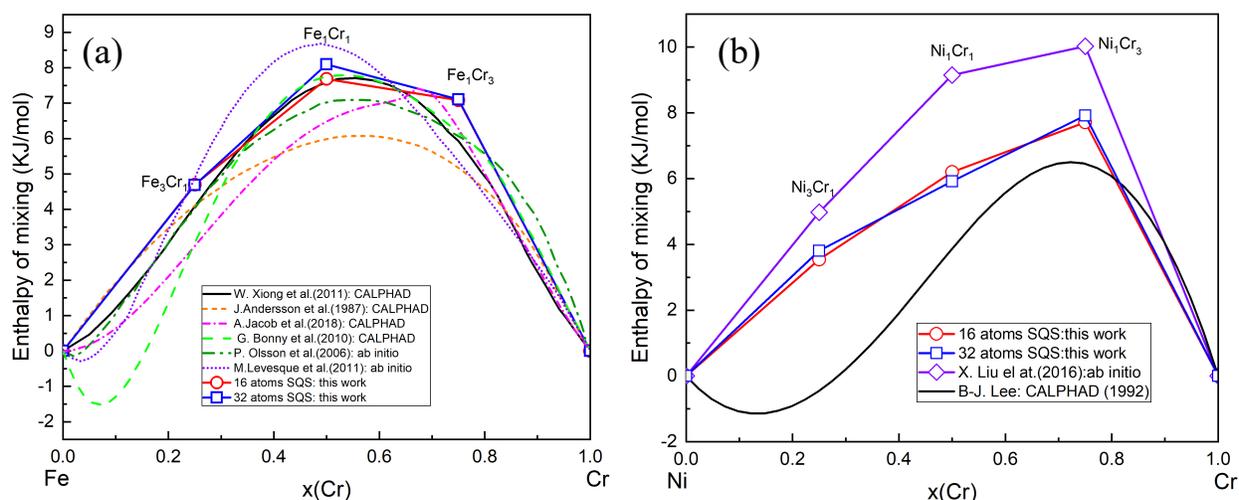

*Figure 3.* $\Delta H_{mix}$ *of BCC phase calculated by ab initio [39-41] and the CALPHAD approach [34-38] for the Fe-Cr (a) and Ni-Cr (b) binary systems*

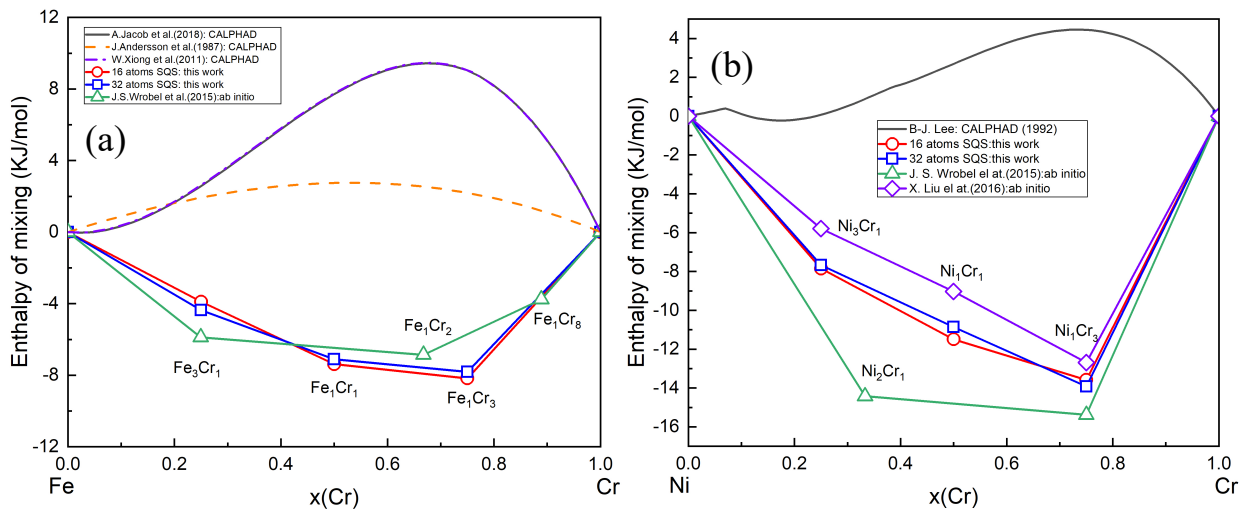

*Figure 4.* $\Delta H_{mix}$ *of FCC phase calculated by ab initio [41, 44] and the CALPHAD approach [34, 35, 37, 38] for the Fe-Cr (a) and Ni-Cr (b) binary systems*

To understand the root of the discrepancy on $\Delta H_{mix}$, the current work investigated the enthalpy of formation ($\Delta H_f$) for these two binary systems. The $\Delta H_f$ of a stable or metastable phase $\Phi$ can be approximated by Eq. (8) [24]:

$$\Delta H_f^\Phi(A_{1-x}B_x) = E^\Phi(A_{1-x}B_x) - (1-x)E^\alpha(A) - xE^\beta(B) \qquad (10)$$

where $E^\Phi(i)$ are the total energy of $A_{1-x}B_x$ in the structure $\Phi$, and pure elements $A$ and $B$ in their stable structure $\alpha$ and $\beta$, respectively. The calculated $\Delta H_f$ of BCC and FCC phases, and their comparisons with CALPHAD are shown in Figure 5. The solid lines with open circles represent the calculated results from the *ab initio* approach, and the dashed lines are the previous CALPHAD results [34, 35, 38]. The blue and green solid lines represent the *ab initio* and CALPHAD FCC baselines, respectively.

It is interesting to observe that the FCC $\Delta H_f$ from *ab initio* and the latest CALPHAD databases are very close to each other on the Fe-rich and Ni-rich side. It is worth noting that the FCC $\Delta H_f$ prediction from the previous two CALPHAD databases [34, 38] are close to each other on the Fe rich side up to Fe₃Cr₁ compositions. Very large differences are observed with higher Cr concentration. The likely explanation is the Gibbs energy description of all the phases other than the FCC phase become much more reliable in the 2018 CALPHAD database [38], which will in turn change the FCC phase Gibbs energy description.

However, significant discrepancies are observed between *ab initio* and CALPHAD results on the Cr-rich side for both systems, which is because the stable/metastable FCC structure becomes unstable with the addition of Cr. In general, the *ab initio* approach has stronger capability to make a reliable prediction on stable/metastable phase than the unstable phase, because full relaxation of unstable phase will break the lattice symmetry [45-47].

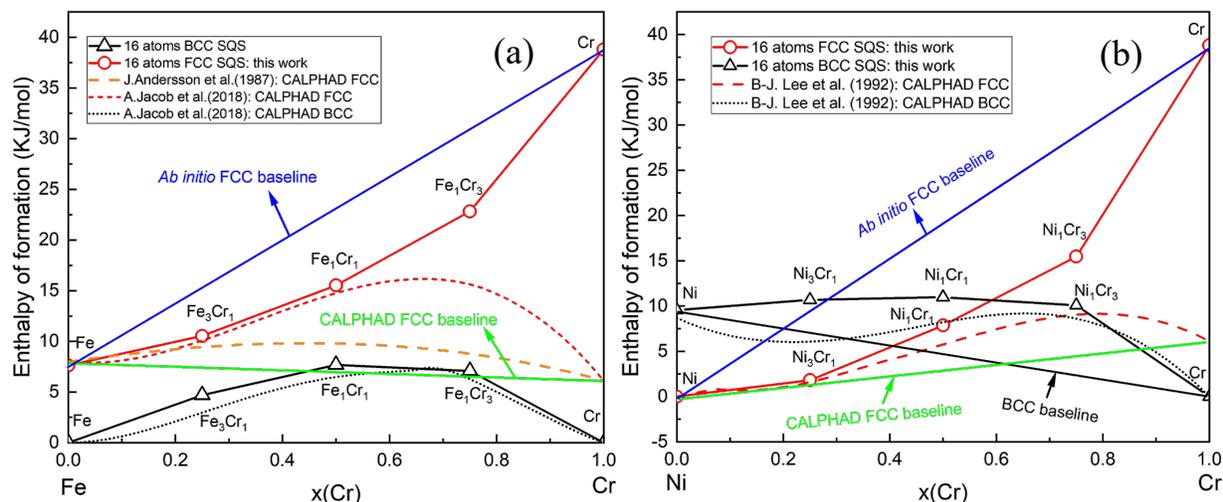

*Figure 5.* $\Delta H_f$ *of FCC and BCC SQS supercells calculated by the ab initio and CALPHAD approaches [34, 38] (a) Fe-Cr (b) Ni-Cr*

After the investigation of $\Delta H_f$, it is necessary to look again the result of $\Delta H_{mix}$. The large differences on $\Delta H_{mix}$ observed in Figure 4 can be explained by the large discrepancy of FCC-Cr lattice stability, which is around 32 kJ/mol. If the *ab initio* Cr lattice stability were chosen as the reference state, it would give a negative $\Delta H_{mix}$ for FCC phase. This is a serious issue, considering the importance of the physical meaning of $\Delta H_{mix}$ in the construction of the Gibbs energy of a solid solution phase. Essentially, the prediction from *ab initio* indicates the attraction of Fe/Ni and Cr atoms to each other in the FCC structure, while the CALPHAD prediction indicates the strong repulsion to each other. The reliability of Cr lattice stability derived by *both approaches* will be systematically discussed in *Section 4*.

### 3.3 Thermodynamic modeling of the FCC phase in the Fe-Cr and Ni-Cr binary systems

In this section, the Fe-Cr and Ni-Cr binary systems were selected as two case studies to verify if the *ab initio* Cr lattice stability can be applied directly into the CALPHAD modeling. The Fe-Cr system was selected as its FCC phase is not stable in most cases and only stable in the Fe rich side in high temperature and form the γ-α Loop. Meanwhile, the Ni-Cr system was selected as its FCC phase is stable and has very large solubility in the Ni-rich side.

The interaction parameters of the FCC phase were evaluated using the Parrot module [48] in Thermo-Calc [49]. This program can take various kinds of experimental data and *ab initio* data concurrently [50]. It works by minimizing the error weighted and summed over each of the selected data. The weight is chosen and adjusted based on the data uncertainties given in the original publications and the judgments by the present authors considering all the SQS data and phase boundary data simultaneously. In order to integrate the *ab initio* Cr lattice stability to the CALPHAD modeling, the current work took the *ab initio* Cr lattice stability at 0K as the enthalpy term, and the entropy term adopted the SGTE value. Specifically, only the FCC interaction parameters were evaluated in both systems. The Gibbs energy description of other phases remains unchanged. The Cr lattice stability obtained by both SGTE and *ab initio* approach and the optimized interaction parameters of FCC phase are listed in Table 2:

*Table 2. Cr lattice stability and FCC Gibbs energy descriptions in the Fe-Cr and Ni-Cr binary systems*

| Cr lattice stability (J/mol) | FCC interaction parameters | References |
|---|---|---|
| $\Delta G^{FCC\text{-}BCC}$=7,284+0.163T | $^{0}L_{Cr,Fe}^{FCC}$= 10,833−7.477T, $^{1}L_{Cr,Fe}^{FCC}$=1,410 | [34] |
| (SGTE) | $^{0}L_{Cr,Fe}^{FCC}$= 28,767−21T, $^{1}L_{Cr,Fe}^{FCC}$=33,057−20.7T | [38] |
|  | $^{0}L_{Cr,Ni}^{FCC}$= 8,030-12.8801T, $^{1}L_{Cr,Ni}^{FCC}$=33,080-16.0362T | [35] |
| $\Delta G^{FCC\text{-}BCC}$=38,800+0.163T | $^{0}L_{Cr,Fe}^{FCC}$= −30,441.6−26.09T, $^{1}L_{Cr,Fe}^{FCC}$=−16,906.5−26.75T | The present |
| (*Ab initio*) | $^{2}L_{Cr,Fe}^{FCC}$=−24,125.7 | work |
|  | $^{0}L_{Cr,Ni}^{FCC}$= −50,807.7−15.5T, $^{1}L_{Cr,Ni}^{FCC}$=−8,203.5−28.28T |  |
|  | $^{2}L_{Cr,Ni}^{FCC}$=−34,928.0 |  |

### 3.3.1 Modeling of the FCC phase in the Fe-Cr phase diagram based on ab initio energy of FCC-Cr

The evaluation of model parameters of the FCC phase in the Fe-Cr binary system started with the fitting of γ-α loop based on the previous database from Jacob et al. [38]. Previously, in CALPHAD, there is no reliable thermochemical data available for the FCC phase. The Gibbs energy of the FCC phase is purely determined by the BCC Gibbs free energy as well as the γ-α loop. Any change of the BCC phase will greatly affect the interaction parameters of the FCC phase. That is why the FCC interaction parameters obtained by Jacob et al. [38] can be very different from the FCC interaction parameters proposed by Andersson et al. [34], as the BCC interaction parameter was updated. Currently, we started to fit the γ-α loop first based on the phase equilibrium data and then considered the *ab initio* $\Delta H_{mix}$ of the FCC phase obtained by SQS calculations. As the *ab initio* thermochemical data is available for FCC phase now, the interaction parameter $^2L_{Cr,Fe}^{FCC}$ was introduced for the parameter optimization. The results giving the best fit of the γ-α loop and *ab initio* $\Delta H_{mix}$ of the FCC phase were adopted in this work.

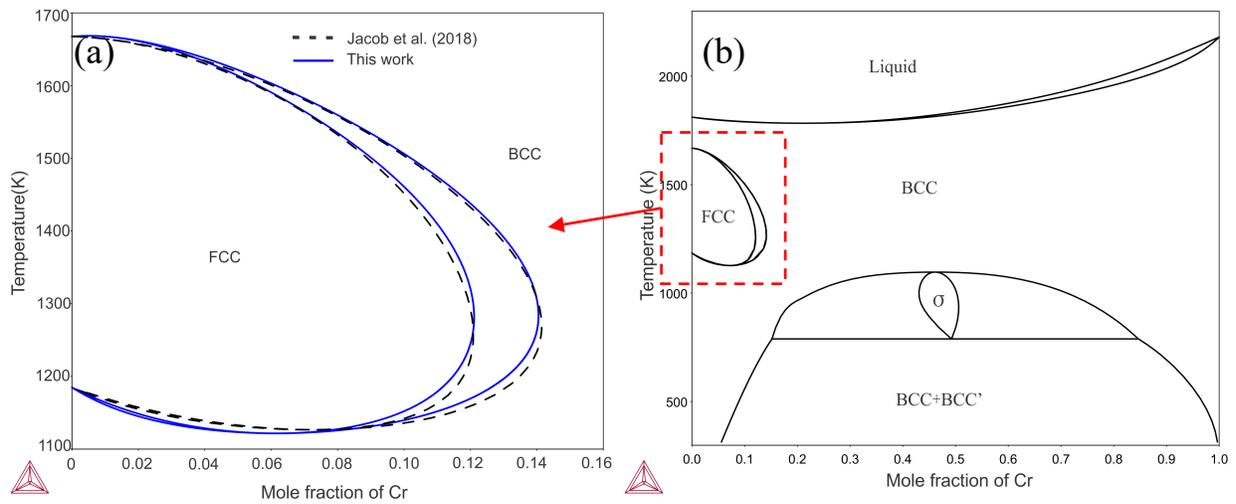

*Figure 6. Fitted γ-α loop based on the ab initio Cr lattice stability (a) γ-α Loop (b) Fe-Cr phase diagram [38]*

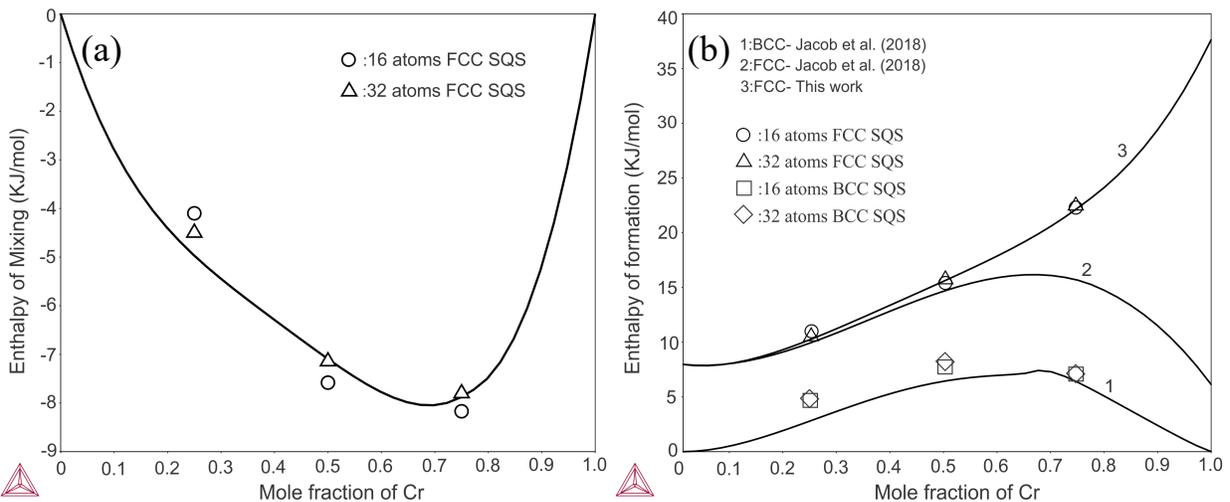

*Figure 7. Optimized thermodynamic descriptions of the FCC phase in Fe-Cr system based on ab initio Cr lattice stability and SQS calculations (a) $\Delta H_{mix}$ (b) $\Delta H_f$*

The updated results from the current work are shown in Figure 6 and Figure 7. Figure 6 shows the updated Fe-Cr binary phase diagram. Especially, Figure 6 (a) shows the excellent agreement on the γ-α Loop with the database from Jacob et al. [38] (black dash lines) and the updated results from the current work (blue solid lines), respectively. The other part of the Fe-Cr phase diagram is identical to the previous version. In addition, Figure 7 shows the $\Delta H_{mix}$ and $\Delta H_f$ of FCC. The enthalpy differences between the optimized results and *ab initio* results for these three compositions ($Fe_3Cr_1$, $Fe_1Cr_1$, and $Fe_1Cr_3$) are less than 1kJ/mol. Consequently, it can be concluded that the *ab initio* FCC-Cr lattice stability at 0K is able to be used successfully in the CALPHAD modeling and reproduce well the γ-α loop of the Fe-Cr phase diagram.

### 3.3.2 Modeling of the Ni-Cr phase diagram based on ab initio energy of FCC-Cr

The Ni-Cr phase diagram based on the *ab initio* Cr lattice stability comparing with the experimental data [51-58], and previous CALPHAD work [35] is shown in Figure 8. It should be noted that the low-temperature intermetallic phase $CrNi_2$ is neglected in the present work due to the complexity of magnetic contribution [59]. In general, the reproduced Ni-Cr phase diagram (blue solid line) shows good agreement with the previous database (black dash line) from Lee et al. [35] as well as the experimental data. In addition, the enthalpy differences between the optimized results and *ab initio* results for these three compositions ($Ni_3Cr_1$, $Ni_1Cr_1$, and $Ni_1Cr_3$) are less than 1kJ/mol as shown in Figure 9. Overall, it is demonstrated that the *ab initio* Cr lattice stability can be successfully applied for the modeling of the Ni-Cr phase diagram.

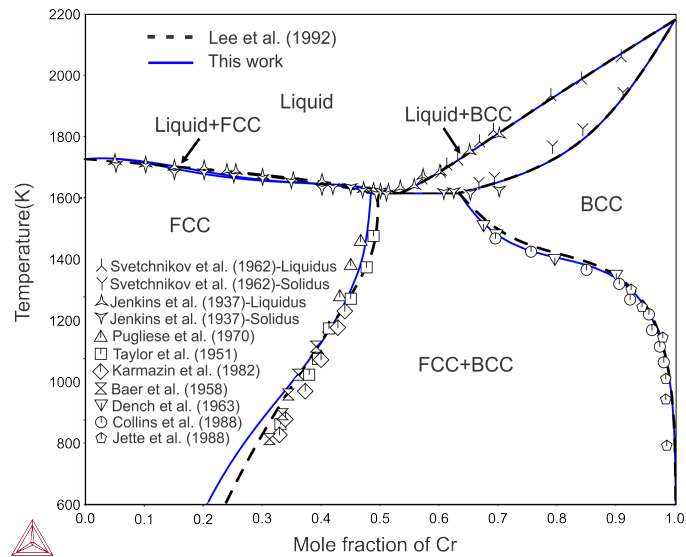

*Figure 8. Reproduced Ni-Cr phase diagram based on the ab initio Cr lattice stability in comparison with experimental data [51-58] and the previous CALPHAD work [35].*

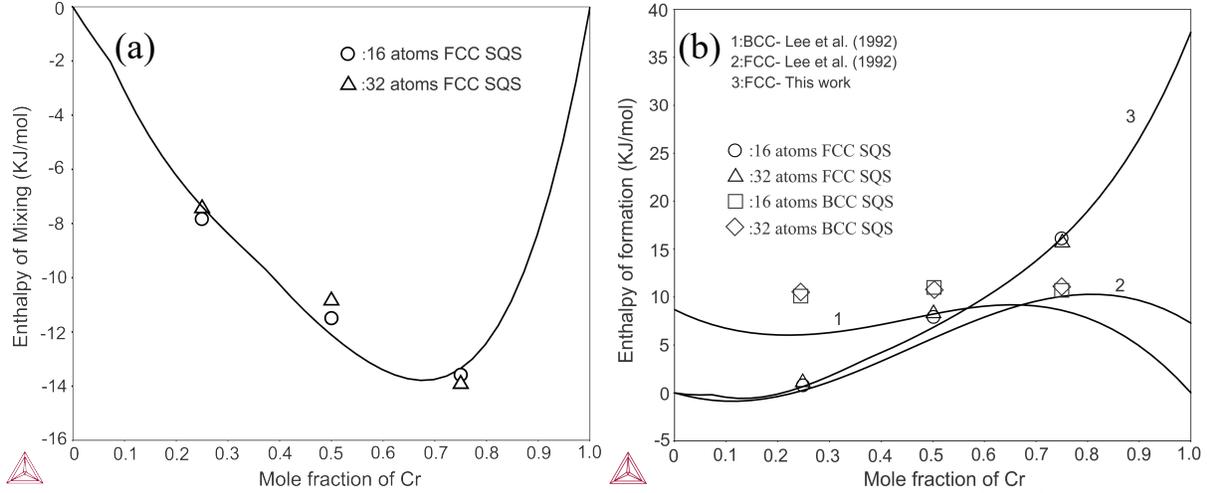

*Figure 9. Optimized thermodynamic descriptions of the FCC phase in Ni-Cr system based on ab initio Cr lattice stability and SQS calculations (a) $\Delta H_{mix}$ (b) $\Delta H_f$*

### 3.3.2 Metastable Equilibrium of Liquid and FCC phase in the Fe-Cr and Ni-Cr binary systems

To understand how the Cr lattice stability will affect the relative stability of the FCC phase and the liquid phase, we investigated the metastable equilibrium phase diagrams for these two systems. The metastable equilibriums of the liquid and FCC phases in the Fe-Cr and Ni-Cr systems are shown in Figure 10 and Figure 11, in which the BCC and σ phases are excluded. According to the previous databases [35, 38] shown in Figure 10 (a) and Figure 11(a), the FCC phase is more stable than the liquid phase at low temperatures in both the Fe-Cr and Ni-Cr systems. It is worth noting that the change of liquidus with respect to the Cr concentration is not monotonical for the Fe-Cr metastable phase diagram (a minimum appears at x(Cr)=0.71), but it is monotonical for the Ni-Cr system. The metastable melting point of FCC-Cr is around *1,475K*.

By contrast, the current work shows that the liquid and FCC phases will coexist at low temperatures when the *ab initio* Cr lattice stability was used as shown in Figure 10 (b). More importantly, the FCC melting temperature will be considered as a negative value on the Cr-rich side of the Fe-Cr and Ni-Cr binary systems, which is due to the large energy difference between the FCC phase based on *ab initio* calculations and the liquid phase. Although the negative melting point could be counterintuitive, it can be interpreted as the amorphous state is more stable than the competing FCC-Cr crystalline configuration at low temperatures, which does not violet the third law as neither FCC nor liquid are stable at low temperatures in comparison with the stable BCC phase [60]. There is actually experimental evidence in the W and Mo, which belong to the same VI-A group in the periodic table as Cr, that the composition region on W-rich and Mo-rich sides of the Cu-Mo and Cu-W binary systems are favored for the formations of amorphous alloy instead of the FCC solution phase [61].

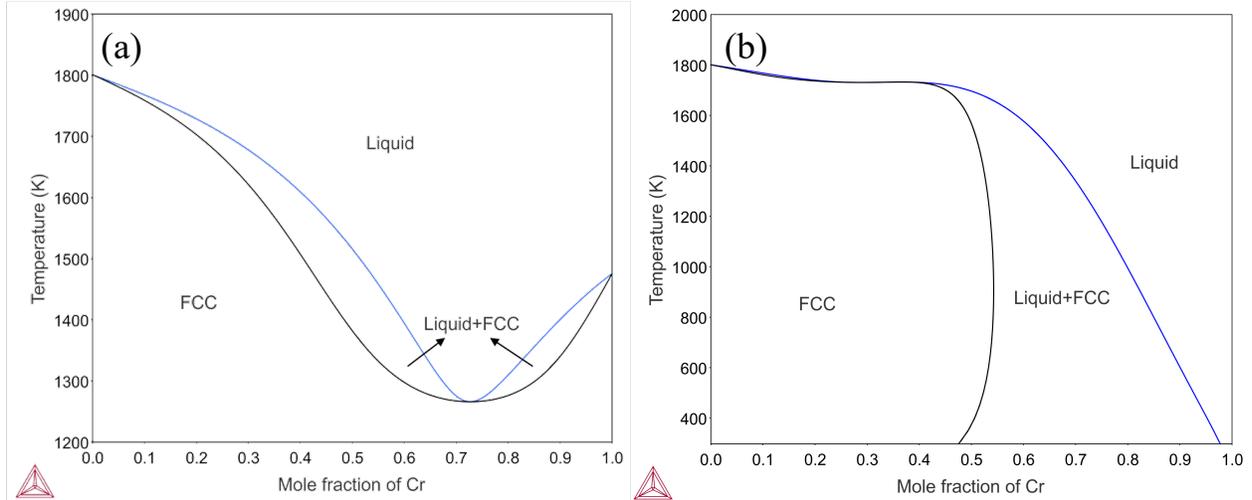

*Figure 10. Metastable equilibrium of liquid and FCC phases in the Fe-Cr binary system with (a) the previous CALPHAD database [38] (b) ab initio Cr lattice stability (this work)*

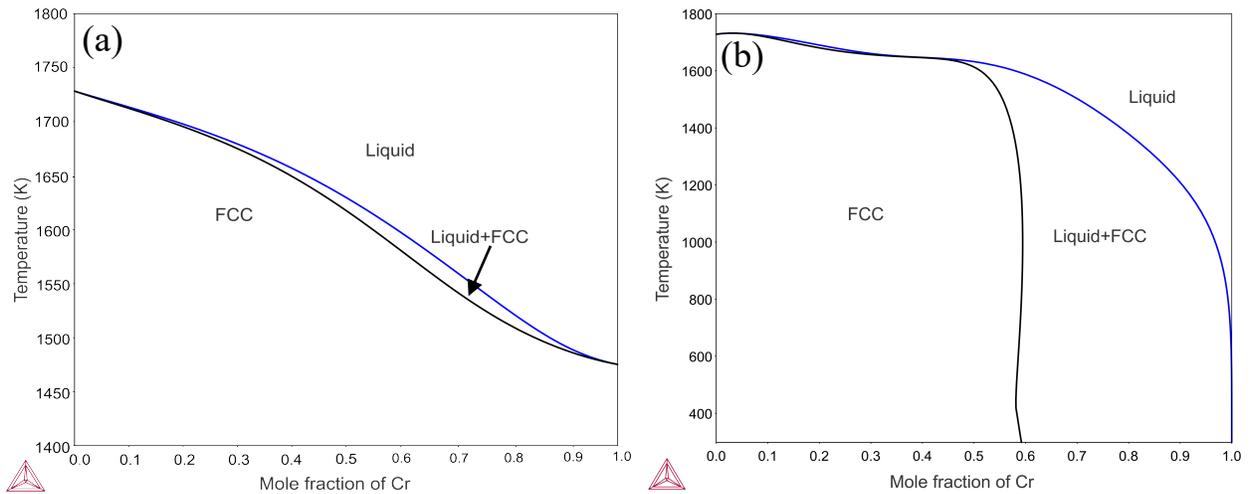

*Figure 11. Metastable equilibrium of liquid and FCC phases in the Ni-Cr binary system with (a) the previous CALPHAD database [35] (b) ab initio Cr lattice stability (this work)*

# 4. Discussion

## 4.1 Discrepancy of Cr lattice stability between ab initio calculations and SGTE

In this section, our discussions are focused on the Cr lattice stabilities among BCC, FCC, and HCP phases ($E^{FCC-BCC}$, $E^{FCC-HCP}$, and $E^{HCP-BCC}$). The comparisons of Cr lattice stability of the current work and other publications [5, 12, 28-31] are shown in Table 3.

*Table 3 Comparison of Cr lattice stability with different groups: (numbers in parentheses are the differences compared with SGTE)*

| Lattice stability (kJ/mol) | This work (DFT) | Wang (DFT) | Material Project (DFT) | OQMD (DFT) | van de Walle (DFT) | SGTE (CALPHAD) | Saunders et al. (CALPHAD) |
|---|---|---|---|---|---|---|---|
| $E^{FCC-BCC}$ | 38.8 (32.67) | 36.8 (30.67) | 37.9 (31.77) | 38.0 (31.87) | 21.4 (15.27)-ID* <br> 37.9 (31.77)-C** | 6.13 | 9.19 (3.06) |
| $E^{FCC-HCP}$ | 1.1 (3.95) | 0.9 (3.75) | 1.3 (4.15) | 0.2 (3.05) | -8.0 (5.15)- ID <br> 0.9 (3.75) -C | -2.85 | -1.82 (1.03) |
| $E^{HCP-BCC}$ | 39.9 (36.62) | 37.7 (34.42) | 39.2 (35.92) | 38.2 (34.92) | 13.4 (10.12)-ID <br> 38.8 (35.52)-C | 3.28 | 7.37 (4.09) |

*: ID- inflection-detection method [12]

**: C-unit cell shape constrains [12]

It can be seen that more than 30 kJ/mol energy differences are observed for $E^{FCC-BCC}$ and $E^{HCP-BCC}$ between *ab initio* calculations at 0K and SGTE for pure Cr lattice stability by the current work and many other publications [5, 12, 28-31]. Recently, van de Walle [12] proposed an inflection-detection (ID) scheme to define the lattice stability of mechanically unstable phases, which reduced the discrepancy of lattice stability between *ab initio* and SGTE to some extent. However, even with this method, we can still observe 15 kJ/mol energy difference for $E^{FCC-BCC}$ and 10 kJ/mol energy difference for $E^{HCP-BCC}$ for pure Cr. Such differences are much larger than what we typically acknowledged between these two approaches, which raises the question: *Should we use the results from CALPHAD or ab initio on the pure Cr lattice stability at 0K*? To answer this question, it is imperative to revisit the history of how the lattice stability of Cr was determined.

### 4.2 Lattice stability of Cr derived by the CALPHAD approach

The concept of lattice stability was first introduced by Kaufman [1] in the 1970s. It was assumed that Cr, Mo, and W have similar properties and should be investigated together as they are in the same VI-A group in the periodic table. The lattice stabilities of Cr and Mo were inferred based on the detailed investigation of W. The melting entropy was assumed to be 2 cal/g-atom for the BCC phase based on experimental melting points of these three elements. The metastable melting point of FCC-W was extrapolated through the extension of the Liquid/FCC trajectory in the W-Ir binary system from the zero-order calculations [1] (ideal solutions, i.e., L=B=E=A=0, *where* L, B, E, and A represent the interaction parameters) as shown in Figure 12. It should be mentioned that Figure 12 is not a regular phase diagram. It is to illustrate the extrapolations of metastable melting temperature from the Liquid/FCC, Liquid/BCC, and Liquid/ HCP trajectory through the zero-order calculations without the consideration of two-phase boundaries. In zero-order calculations, all the interaction parameters were considered to be zero, and the phase boundaries depended only on the free energy differences between BCC, HCP, FCC, and the liquid phase of these pure elements. The metastable melting temperature of FCC-W was estimated to be 2,230K, as shown in Figure 12. The entropy of the melting of the metastable FCC was assumed to be slightly higher than BCC, i.e., 2.15 cal/g-atom. The Gibbs free energy difference between BCC and FCC was estimated as *2,500+0.15T (cal/g-atom)*, which was directly adopted for the other two elements, i.e., Cr and Mo. With that, the metastable melting point of FCC-Cr was determined to be *860K* based on $\Delta G^{L-FCC}$. The detailed values of these three elements are presented in Table 4.

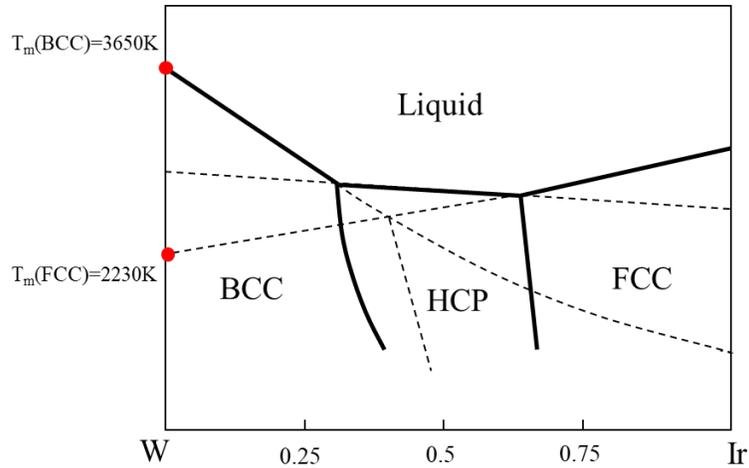

*Figure 12 Illustration of the ideal solution phase diagram (Zero-order calculations)*

*Table 4 Lattice stabilities reproduced from Larry Kaufman's work [1]*

|  | Cr (cal/g-atom) | Mo (cal/g-atom) | W (cal/g-atom) |
|---|---|---|---|
| $\Delta G^{L\text{-}BCC}$ | 4,350-2T | 5,800-2T | 7,300-2T |
| $\Delta G^{L\text{-}FCC}$ | 1,850-2.15T | 3,300-2.15T | 4,800-2.15T |
| $\Delta G^{FCC\text{-}BCC}$ | 2,500+0.15T | 2,500+0.15T | 2,500+0.15T |
| $T_m(BCC)$-K | 2,175 | 2,900 | 3,650 |
| $T_m(FCC)$-K | 860 | 1,530 | 2,230 |

Later, Saunders et al. [28] estimated the lattice stabilities with the same general approach but using improved experimental values of melting entropy and melting points of the BCC phase. the melting entropy was treated separately for these three elements because there is a general tendency that the melting entropy is higher when the melting point is higher [62]. The metastable melting point of FCC-Cr was obtained as *1,350K* through the liquidus extrapolations in the Ni-Cr binary system, where the phase has a reasonable range of FCC existence. In addition, the metastable melting entropies have been estimated assuming they follow the same trends which are observed in stable phases. Finally, the lattice stability of pure Cr was determined to be *9,192-0.89T (J/mol-K)*. The result from Saunders et al. [28] is shown in Table 5. Compared with Kaufman's work, Saunders et al. observed that the disagreement can, in some cases, be decreased if more recent values of the melting entropies are used in the CALPHAD method.

*Table 5 Lattice stabilities according to Saunders et al. [28]*

|  | Cr (J/mol-K) | Mo (J/mol-K) | W (J/mol-K) |
|---|---|---|---|
| $\Delta G^{L\text{-}BCC}$ | 21,004-9.637T | 37,480-12.942T | 52,314-14.158T |
| $\Delta G^{L\text{-}FCC}$ | 11,812-8.747T | 7,680-8.992T | 17,314-9.958T |
| $\Delta G^{FCC-BCC}$ | 9,192-0.89T | 29,800-3.95T | 35,000-4.20T |
| $T_m(BCC)$-K | 2,180 | 2,846 | 3,694 |
| $T_m(FCC)$-K | 1,350 | 854 | 1,739 |

In 1980s, SGTE [63] recommended a new method regarding the extrapolation of the properties of the liquid phase below the melting point and the solid properties above the melting point, which keeps the difference in heat capacity ($C_p$) the same as it is at melting point. In order to apply this recommendation, Andersson et al. [62] revise the lattice stability of the liquid, and such revisions were made in connection with the assessments of numbers of binary systems with Cr, Mo, and W as shown in Table 6. For the metastable FCC-Cr, the melting point and the melting entropy were taken from an early version reported by Saunders and Miodownik [62]. Finally, the lattice stability of pure Cr was determined to be *7,284+0.163T (J/mol-K),* and the metastable melting point of FCC-Cr is chosen as *1,475K.*

*Table 6 Lattice stabilities according to Andersson et al. [62]*

| | Cr (J/mol-K) | Mo (J/mol-K) | W (J/mol-K) |
|---|---|---|---|
| $\Delta G^{L-BCC}$ (T<2180K) | $24,335.93-11.42T+2.3765*10^{-21}T^7$ | $41,616.77-14.61T+4.0358*10^{-22}T^7$ | $52,160.58-14.11T-2.7135*10^{-24}T^7$ |
| $\Delta G^{L-BCC}$ (T>2180K) | $18,405.93-8.56T+2.8853*10^{32}T^{-9}$ | $34,262.10-11.94T+4.6104*10^{33}T^{-9}$ | $52,432.76-14.19T-1.9217*10^{33}T^{-9}$ |
| $\Delta G^{FCC-BCC}$ | 7,284+0.163T | 15,200-0.63T | 19,300-0.63T |
| $T_m(BCC)$-K | 2,180 | 2,896 | 3,695 |
| $T_m(FCC)$-K | 1,475 | 1,735 | 2,229 |

The lattice stabilities derived from Andersson are what SGTE used to describe the Gibbs energy of pure elements in the last 30 years. Figure 13 shows how the lattice stability of pure Cr was derived by the CALPHAD approach and the difference of pure Cr lattice stability between CALPHAD and *ab initio* calculations. In the CALPHAD approach as shown in Figure 2, $T_m^{BCC}$, $G_{BCC}$ (T<2,180K), and $G_L$ (T>2,180K) can be determined experimentally. Meanwhile, $G_{BCC}^{metastable}$ (T>2,180K) and $G_L^{metastable}$ (T<2,180K) are extrapolated, which make use of a special term to make the $C_p$ value of the extrapolated phase approach the $C_p$ of the stable phase gradually [63]. As FCC-Cr is unstable, the melting point of FCC-Cr was estimated by extrapolation in the Ni-Cr binary phase diagram. Its melting entropy was assumed to be close to the melting entropy of BCC-Cr.

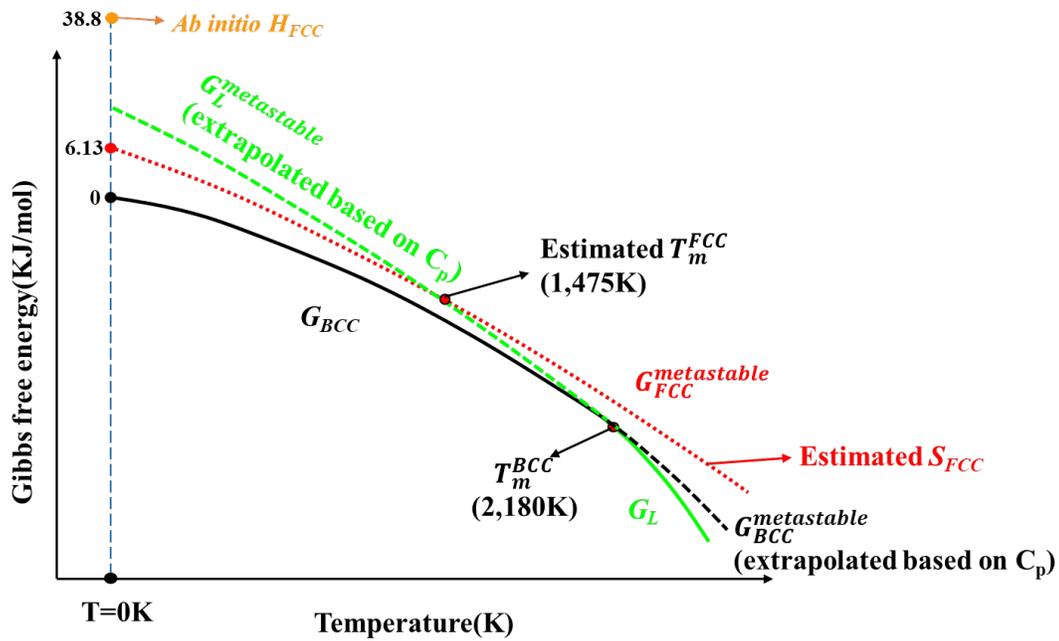

*Figure 13 Lattice stability of pure Cr derived by traditional CALPHAD approach*

However, the above extrapolations are based on three assumptions:

(1) FCC-Cr has a positive metastable melting point.

(2) The metastable melting point of FCC-Cr can be extrapolated by the liquidus trend of various Cr-based binary systems.

(3) The melting entropy of metastable FCC-Cr is very close to that of BCC-Cr.

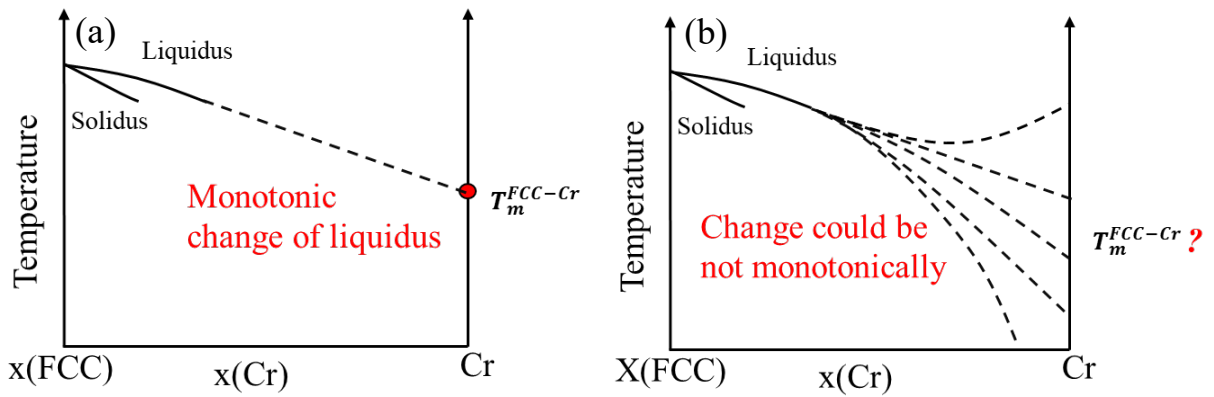

*Figure 14 Schematical extrapolation of metastable/unstable melting point of Cr through the trajectory of liquid/FCC equilibrium in Cr-X(FCC) binary system (a) CALPHAD approach assuming monotonic change of liquidus (b) possible non-monotonic change of liquidus.*

These assumptions were criticized by Hillert et al.[6]. First, it is easy to assume a solid has a positive melting point, but this argument could not apply to all the metastable/unstable phases. Second, the estimation of the FCC-Cr melting point requires the liquidus of the FCC phase in binary systems changes monotonically in the composition range of the extrapolation as shown in Figure 14 (a). However, this type of estimation fails to handle the cases if the change of liquidus is not monotonic in the extrapolated part as schematically shown in Figure 14 (b). Indeed, our previous calculations (*Section 3.3.2*) clearly show that the change of liquid/FCC equilibria is not monotonical for the Fe-Cr binary system (Figure 10 (a)). Third, the melting entropy of FCC-Cr could be very different from the melting entropy of BCC-Cr, because there is a general tendency that the melting entropy is higher when the melting point is higher [62]. Therefore, the lattice stability description for pure Cr derived from the CALPHAD method is problematic [6-8, 13, 60, 62], even though the lattice stability of Cr derived by SGTE has been successfully used in many systems.

### 4.3 Lattice stability of Cr derived by ab initio approach

The *ab initio* calculations utilized methods that solve the Schrödinger equation and start with the specification of the lattice space group taking into account a wide variety of interatomic forces. It is, therefore, possible to make a valid prediction on the lattice stabilities of elements at the ground state with different crystal structures. However, it has been argued that the *ab initio* approach has limited capability to predict the lattice stability of unstable phases [5, 7, 64-66]. Many studies [7, 64, 65] observed the imaginary frequencies of phonon modes for unstable structures, which made people question the *ab initio* results since the atoms would no longer occupy an ideal lattice position with tiny lattice vibrations [5, 7, 64-66]. The current work, as shown in Figure 15, also observed the imaginary frequencies for FCC-Cr, which means that it is unstable.

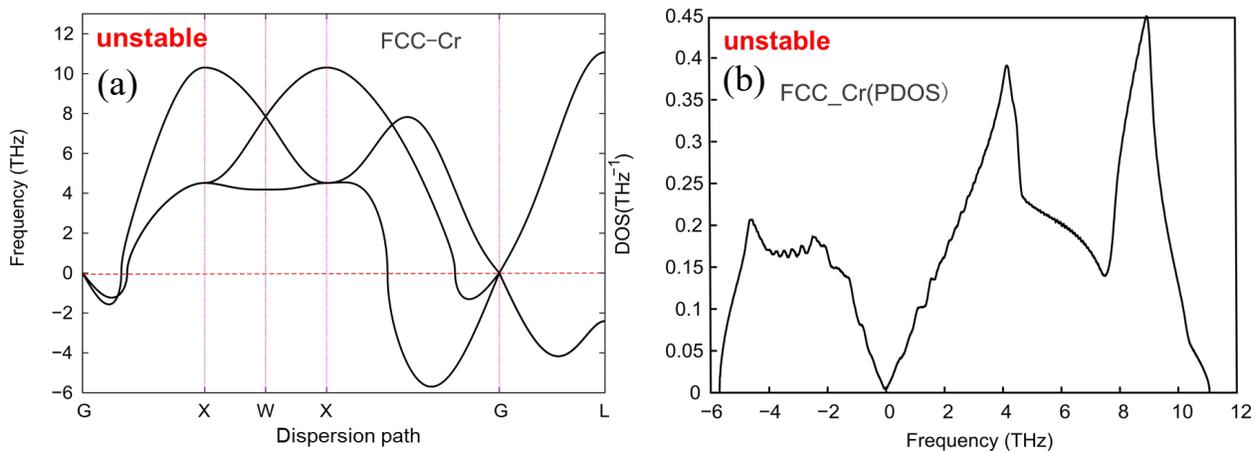

*Figure 15 Phonon properties of FCC-Cr (a) Phonon dispersion (b) Phonon density of state*

To further investigate the instability behavior of FCC-Cr, one technique is to study the tetragonal transformation path. The lattice instability along the tetragonal transformation path of the cubic metals represents the transformation between BCC (c/a=1) and FCC (c/a=1.414) structures, which was studied by many research groups [5, 67-71]. The tetragonal transformation path of Cr and Fe is demonstrated in Figure 16. It can be observed that there is no inflection point during the transformation of BCC-Fe to FCC-Fe, which means FCC-Fe exhibits no instability during the tetragonal transformation [11, 12]. Accordingly, not much discrepancy is observed for the lattice stability of Fe between *ab initio* calculation and SGTE. However, the total energy of FCC-Cr is a local maximum in the tetragonal transformation process, which indicates that it is unstable. It has been reported [5] that the higher the local maximum is, the larger the discrepancy between the *ab initio* results and SGTE data, but the reason for this discrepancy has not been examined in detail.

Recently, van der Walle et al. [11] proposed an inflection-detection method to handle the instabilities of crystal structures. They postulated that the correct total energy of the unstable phase is around the inflection point in Figure 16 along a transformation path as that point is considered as the threshold in which the harmonic description of the vibrational entropy can be interpreted correctly. However, there are some issues with this approach. First, the crystal structure of Cr at the inflection point (c/a=1.2) departs substantially from the ideal FCC structure (c/a=1.414), so it does not represent the total energy of a rigid FCC or FCC-like structure. Second, this method does not apply to all the elements. For elements such as Al, Pt, and Tc, the lattice stabilities calculated by the inflection-detection method exhibit much larger discrepancies with SGTE compared with the cell-shape constrained methods [12]. Even for Cr, the energy difference ($\Delta E^{FCC-BCC}$) between the inflection-detection method and SGTE is as large as 15.27 kJ/mol, which is still significantly larger than the acceptable error range. Although the discrepancy of lattice stability between SGTE and *ab initio* can be reduced to some extent by using the inflection-detection method, it still cannot be considered as a practical solution to reconcile these two approaches.

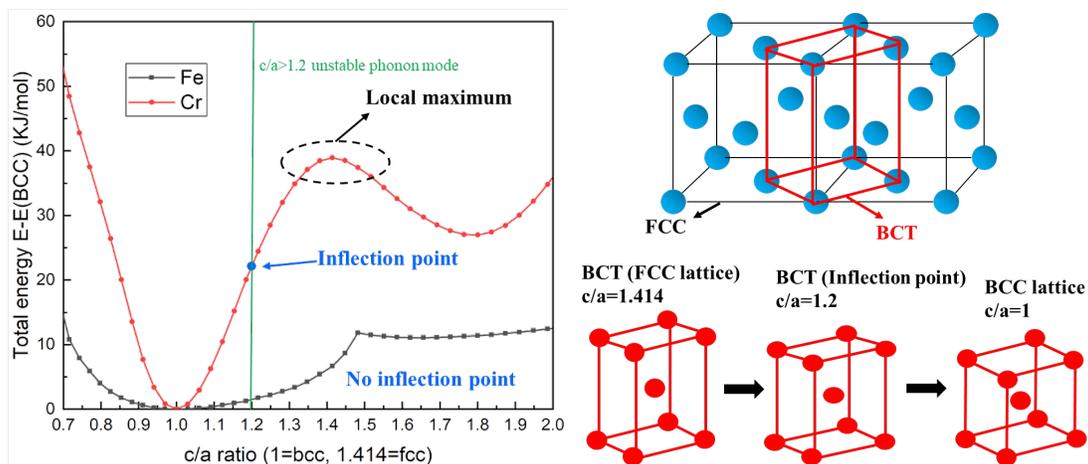

*Figure 16 Total energy E-EBCC, along the tetragonal transformation path for Fe and Cr*

Although one can use a harmonic expansion of the structure near the inflection point and ignore the unstable phonon mode, it does not represent the real total energy of the rigid FCC or FCC-like lattice. The tetragonal transformation path shows that the total energy increase with the increase of *c/a*. As the FCC *c/a* is much larger than the one at the inflection point (*c/a=1.2*), it seems that the correct energy should be higher even though the imaginary frequency is observed. Accordingly, we postulate that the total energy of the FCC-Cr lattice should be much higher than the value near the inflection point proposed by van der Walle et al. [11, 12]. To further understand the reliability of FCC-Cr lattice stability, we should check its mathematical expression. The lattice stability of FCC-Cr can be represented by a simple expression on $H_{FCC}^{Cr} - TS_{FCC}^{Cr}$, where $H_{FCC}^{Cr}$ (enthalpy) and $S_{FCC}^{Cr}$ (vibration entropy) are related to the total energy and phonon density of state (PDOS), respectively. For FCC-Cr, the harmonic description of its vibration entropy is thermodynamically incorrect because of the imaginary frequency and the $S_{FCC}^{Cr}$ cannot be physically defined, but should be considerably larger than the stable structures due to entropy divergence near instability [9, 10]. However, for $H_{FCC}^{Cr}$, researchers have divided opinions about its reliability. Saunders et al. [7, 8, 60] reported that the *ab initio* calculations have the inherent capability of yielding accurate energy value for metastable FCC structure at 0K. which is also supported by Grimvall et al. [7, 8]. He claimed that $H_{FCC}^{Cr}$ has a well-defined physical meaning if it is calculated by the rigid structure. By contrast, Asker et al. [66] illustrated that *ab initio* calculations at 0 K of the static lattice could not provide accurate information about the lattice stability because of the instabilities. Even though we cannot claim that the current value for $H_{FCC}^{Cr}$ is error-free, it is safe to claim that the exact $H_{FCC}^{Cr}$ should be very different from the SGTE result and much higher than the result from the inflection-detection method proposed by van der Walle et al. [11].

### 4.4 CALPHAD vs. ab initio Cr lattice stability

Based on the discussion above, it is evident that neither the CALPHAD nor the *ab initio* approach could provide a reliable value for the lattice stability of FCC-Cr. Accordingly, which $H_{FCC}^{Cr}$ should be adopted is still a choice. The current section will focus on the discussion of the *pros and cons* for the adoption of each Cr lattice stability.

The SGTE Cr lattice stability has been successfully implemented in all CALPHAD databases developed so far. They are based on the assumption of a small deviation from the regular or subregular solution behavior, which has the strong empirical capability to expand the model for the prediction of high-ordered systems. By contrast, the *ab initio* calculations have the inherent capability of yielding an accurate value of any given structures at 0K, but have little or no capability of predicting the temperature dependence of Gibbs energy in cases where instabilities are involved, such as the FCC-Cr [60] discussed in the current work. In general, researchers are fully relying on the *ab initio* calculations for the low-temperature thermodynamic properties of stable structures.

The integration of *ab initio* calculations with the CALPHAD modeling is the trend for the database assessment [10, 24]. However, the discrepancies between *ab initio* and SGTE lattice stabilities largely impede the full integration of *both approaches*. If we continue to use the SGTE lattice stabilities, there are some critical issues. Similar to the discussion on the Cr lattice stabilities, the *ab initio* $\Delta H_{mix}$ cannot be applied to construct the Gibbs free energy for the systems with lattice stability difference over 10kJ/mol, including V, Nb, Mo, Tc, Ru, Rh, Pd, W, Re, Os, and Ir. Such an issue could be solved if the *ab initio* lattice stability is adopted. Moreover, the same problem is also existing for the multi-component solution phases. In the CALPHAD approach, sublattice models were adopted, typically with end members of pure elements. The lattice stability of the unstable/metastable endmembers was traditionally chosen as an arbitrary 5kJ/mol higher than the pure elements' stable structure. However, this estimation is physically not correct as different solution phases should have different end-member energies [72-75]. According to the aforementioned arguments, the present work demonstrated that it is promising to construct a new CALPHAD database purely based on *ab initio* data. The eventual solution would be developing the thermodynamic database by using software such as PyCalphad [76] and Extensible Self-optimizing Phase Equilibria Infrastructure (ESPEI) [77], which can automatically generate and optimize the interaction parameters based on the new data achieved from *ab initio*, particularly the high throughput calculations [78], machine learning [79], or experiments, particularly the high throughput approaches [80].

# 5. Conclusions

In the current work, we systematically investigated the lattice stability of Cr from both the *ab initio* and CALPHAD approaches. In addition, the Fe-Cr and Ni-Cr binary systems were chosen as case studies to verify the capability of the CALPHAD treatment of *ab initio* Cr lattice stability. The enthalpy of mixing ($\Delta H_{mix}$) and enthalpy of formation ($\Delta H_f$) of both FCC and BCC of the Fe-Cr and Ni-Cr binary systems were studied. After the comprehensive discussion on the reliability of the Cr lattice stability, it is demonstrated that neither the CALPHAD nor the *ab initio* approach could provide a reliable value for the lattice stability of FCC-Cr. The main issue for the CALPHAD approach is that the Gibbs energy of fcc-Cr is based on empirical extrapolations and assumptions. Meanwhile, we cannot claim that the *ab initio* $H_{\text{FCC-Cr}}$ is error-free, However, the current work shows that the *ab initio* $H_{\text{FCC-Cr}}$ would be a better scientific guess. It is observed that the significant differences of FCC $\Delta H^{mix}$ between the *ab initio* and CALPHAD approaches in these two binary systems are attributed to the discrepancy of pure Cr lattice stability. The present work shows that the *ab initio* Cr lattice stability can be successfully applied to the CALPHAD modeling and reproduce the Fe-Cr and Ni-Cr phase diagrams. Furthermore, the present work proved the concept on how to reconcile the lattice stability discrepancy issue for Cr, similar work can be carried out for other elements and systems, including the greatly debated Mo-Ru system [81]. Consequently, the adoption of *ab initio* lattice stability even for the unstable structures has the potential to dramatically accelerate the database development in the future by using software such as PyCalphad and ESPEI.

# 6. Acknowledgement and Disclaimer

This material is based upon work supported by the Department of Energy under award number **DE-FE0030585** (SY and YZ) and National Science Foundation under award number **CMMI-1825538** (YW and ZKL). The authors would like to thank the support and guidance from the DOE National Energy Technology Laboratory program manager, Maria M. Reidpath. This paper was prepared as an account of work sponsored by an agency of the United States Government. Neither the United States Government nor any agency thereof, nor any of their employees, makes any warranty, express or implied, or assumes any legal liability or responsibility for the accuracy, completeness, or usefulness of any information, apparatus, product, or process disclosed, or represents that its use would not infringe privately owned rights. Reference herein to